\documentclass[10pt,twocolumn,twoside]{IEEEtran}
\usepackage{blindtext, graphicx}
\usepackage[utf8]{inputenc}
\usepackage{listings}
\usepackage{amssymb}
\usepackage{amsmath}
\usepackage{mathtools}
\usepackage{algorithm} 
\usepackage{algpseudocode} 
\usepackage{color}
\usepackage{caption}
\usepackage{subcaption}
\usepackage{enumitem}
\usepackage{tabularx}
\usepackage[table]{xcolor}
\usepackage{cite}
\usepackage{caption}
\usepackage{xcolor}
\usepackage{soul}
\usepackage{booktabs}
\usepackage{makecell}

\usepackage[compact]{titlesec}
    \titlespacing{\section}{0pt}{1.0ex}{0.4ex}
    \titlespacing{\subsection}{0pt}{1.45ex}{0.45ex}
    \titlespacing{\subsubsection}{0pt}{1.0ex}{0.45ex}

\algnewcommand{\Initialize}[1]{%
  \State \textbf{Initialize:}
  \Statex \hspace*{\algorithmicindent}\parbox[t]{.8\linewidth}{\raggedright #1}
}
\algnewcommand{\Input}[1]{%
  \State \textbf{Input:}
  \Statex \hspace*{\algorithmicindent}\parbox[t]{.8\linewidth}{\raggedright #1}
}

\begin{document}

\title{
Deep Reinforcement Learning Based Placement for Integrated Access Backhauling in UAV-Assisted Wireless Networks
}



\author{Yuhui Wang,~\IEEEmembership{Student Member,~IEEE} and Junaid Farooq,~\IEEEmembership{Member,~IEEE}
\thanks{\hrule \vspace{0.1in}Yuhui Wang and Junaid Farooq are with the Department of Electrical \& Computer Engineering, College of Engineering and Computer Science, University of Michigan-Dearborn, Dearborn, MI 48128 USA. E-mails: \{ywangdq, mjfarooq\}@umich.edu.\\
\indent Preliminary results from this work have appeared in Proceedings of the $20^{\text{th}}$ IEEE International Conference on Mobile Ad-Hoc and Smart Systems (MASS), International Workshop on Unmanned Autonomous Vehicles and IoT (UAV-IoT 2023)~\cite{yuhi_ieee_mass}.
}
}


\maketitle

\begin{abstract}
The advent of fifth generation (5G) networks has opened new avenues for enhancing connectivity, particularly in challenging environments like remote areas or disaster-struck regions. Unmanned aerial vehicles (UAVs) have been identified as a versatile tool in this context, particularly for improving network performance through the Integrated access and backhaul (IAB) feature of 5G. However, existing approaches to UAV-assisted network enhancement face limitations in dynamically adapting to varying user locations and network demands. This paper introduces a novel approach leveraging deep reinforcement learning (DRL) to optimize UAV placement in real-time, dynamically adjusting to changing network conditions and user requirements. Our method focuses on the intricate balance between fronthaul and backhaul links, a critical aspect often overlooked in current solutions. The unique contribution of this work lies in its ability to autonomously position UAVs in a way that not only ensures robust connectivity to ground users but also maintains seamless integration with central network infrastructure. Through various simulated scenarios, we demonstrate how our approach effectively addresses these challenges, enhancing coverage and network performance in critical areas. This research fills a significant gap in UAV-assisted 5G networks, providing a scalable and adaptive solution for future mobile networks.
\end{abstract}


\IEEEpeerreviewmaketitle

\section{Introduction}


The advent of fifth-generation (5G) cellular networks has ushered in a wealth of opportunities for enhancing connectivity and coverage, particularly in remote or rural areas with limited network infrastructure. Unmanned aerial vehicles (UAVs) have emerged as a promising tool to augment the performance of these networks, capitalizing on their versatility and mobility, especially for mission-critical operations \cite{9771888,9839646}. In the context of UAV-assisted 5G network, the UAV is connected to a base station over the radio access network (RAN), which is strategically divided into two distinct units: the centralized unit (CU) and the distributed unit (DU) \cite{9977571}. This division aims to optimize network efficiency and enhance overall performance by allocating specific functions to each unit. The DU, a crucial component of this split architecture, can be conveniently positioned on UAVs, enabling them to serve as integrated access and backhaul (IAB) nodes. This empowers the UAVs to act as dynamic, mobile base stations, enhancing network coverage and connectivity. On the other hand, the CU is typically situated at the traditional base stations (BS), housing the centralized processing and control functions. This segregation of responsibilities allows for a more distributed and flexible network architecture, enabling UAVs to efficiently handle communication tasks at the edge of the network, while the centralized units manage core functionalities and resource allocation \cite{9857929}. This paper addresses the challenge of optimizing end-to-end performance in UAV-assisted cellular networks, focusing on the dynamic and adaptive 3D placement of UAVs. 

\begin{figure}[t!]
    \centering
    \includegraphics[width=\linewidth]{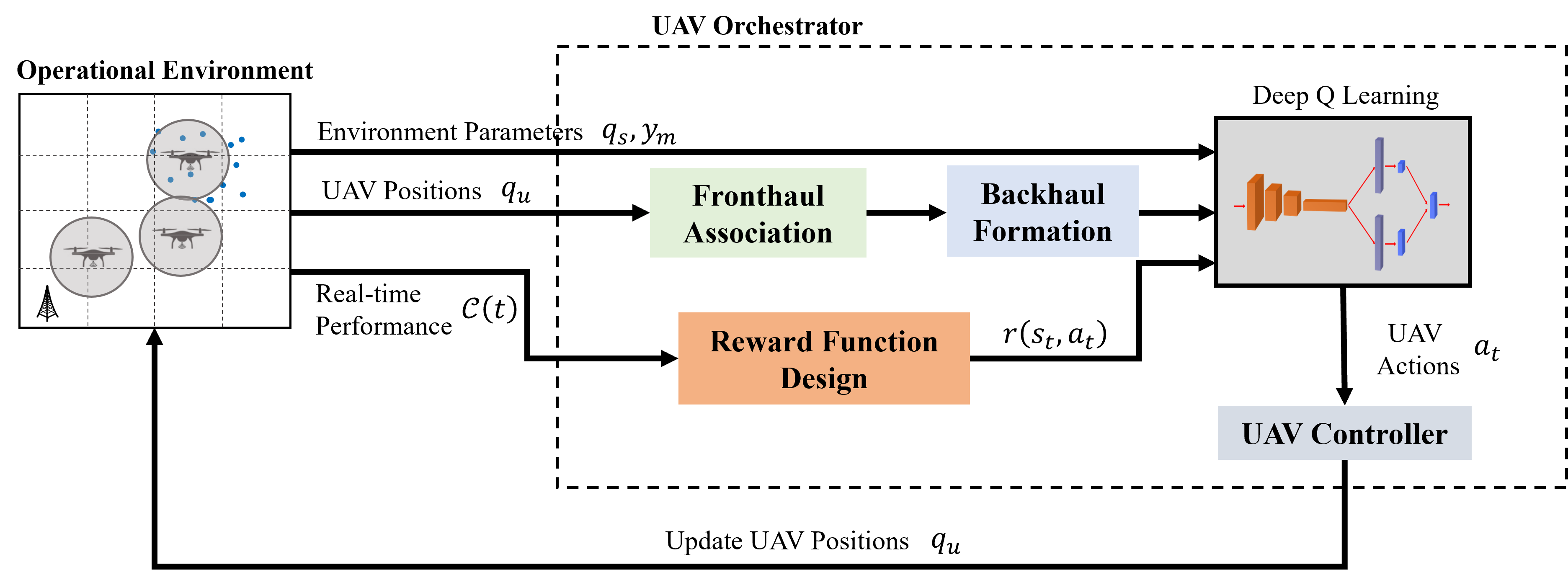}
    \caption{Overview of the deep reinforcement learning based placement optimization framework for integrated access backhauling in UAV networks. \vspace{-0.0in}}
    \label{fig:notation}
\end{figure}

The fluid nature of the network topology, characterized by coverage gaps and rapidly changing user locations, poses significant challenges for ensuring consistent connectivity and optimal performance. Integrated access and backhaul in 5G presents a potential solution that can utilize UAVs to augment coverage for short durations in critical areas \cite{sookhak2022joint}. However, the demand for enhanced coverage can unpredictably arise at various locations and times, making conventional fixed placement strategies impractical and ineffective \cite{9716752}. Furthermore, an essential consideration in UAV deployment is the balance between maintaining robust backhaul connectivity, facilitating seamless data transfer between UAVs and the central infrastructure, and optimizing fronthaul communication with ground users. This tradeoff requires adaptive and dynamic decision making as the prioritization of one link can potentially compromise the performance of the others \cite{dabiri20223d}. For instance, in natural disasters such as a hurricane or earthquake, where traditional terrestrial network infrastructure has been severely impacted, cellular users may be able to connect to a mobile BS mounted on a UAV. Since, the UAV has a limited communication range, it may require multi-hop links over neighbouring UAVs to connect to the nearest operational cellular tower. While the access link between UAV and the users is critical, it is also crucial to maintain the backhaul multi-hop links for successful end-to-end communication. This requires careful orchestration of the UAV network leveraging artificial intelligence (AI)-enabled decision-making and placement optimization techniques~\cite{UAV_assistance}.

To tackle these challenges, we propose a dynamic, real-time approach based on deep reinforcement learning (DRL). Fig.~\ref{fig:notation} shows framework of the UAV placement algorithm for optimal end-to-end performance. This approach is designed to be adaptive, capable of learning the optimal UAV placement on the fly and continuously updating it based on real-time performance measurements. This is particularly crucial for critical users who may lack sufficient connectivity for mission-critical performance needs and whose locations can significantly fluctuate over time. In contrast to the existing works in the literature that have various objectives around optimization \cite{huang2022deployment, karmakar2022reliable}, we consider an objective based on balancing the tradeoff between backhaul and fronthaul links. The proposed approach integrates a data rate-aware reward function with a reinforcement learning algorithm, which enables the UAV to learn and adapt to its environment by optimizing its placement based on feedback. Additionally, the data rate-aware reward function ensures that the UAV prioritizes the strength of the wireless links while optimizing its placement, resulting in an improved overall performance of the system.


The rest of the paper is organized as follows: Section II provides an overview of the related works in literature, Section III describes the DRL-based system framework for UAV placement, Section IV formulate the optimization problem and present the details of the algorithm, Section V provides the simulation results and performance evaluation, and finally Section VI concludes the paper.

\section{Related Work}

The use of UAVs in wireless networks has been a topic of significant interest, with potential applications as aerial base stations (BSs) in 5G networks \cite{khan2020uav,9739009,10299642}, mobile edge-cloud computing systems \cite{9858158,9771416,zhou2022uav}, scalable nodes in ad hoc networks \cite{9700761,bansal2022shots,abdulhae2022cluster}, and data collection and surveillance \cite{9682599,wang2022zero,9864220}. Key challenges in this field include optimizing UAV placement, trajectory control, and maintaining quality of service (QoS) for ground users. Various solutions have been proposed, including centralized optimization, distributed control-based algorithms, and learning-based methods.

\subsection{UAV Trajectory Design}
UAV trajectory design aims to determine the optimal routes and motion patterns for UAVs to achieve specific objectives, such as data collection and surveillance. It is an essential aspect for ensuring physical layer security \cite{na2022joint}, optimizing power allocation \cite{wu2022trajectory} and avoiding collision with obstacles \cite{wang2022learning}. In \cite{zargari2022user}, the authors investigate the trajectory optimization problem for a UAV equipped with intelligent reflecting surface (IRS). In order to improve users uplink signal and the network energy efficiency, they propose a solution based on the alternating optimization algorithm. Results show the reduction in the cost of circuit power consumption as well as improvement in uplink data transmissions.
In \cite{cui2020adaptive}, the authors propose a model-free solution to optimize UAV trajectories while ensuring QoS for ground users. Their approach uses an adaptive reinforcement learning algorithm that learns the optimal trajectory of the UAV based on a QoS-aware reward function. Similarly, Guo et al. focus on dynamic entry and exit for balanced offloading in IoT networks~\cite{guo2022achieve}. They propose a trajectory optimization scheme that maximizes the energy efficiency of the UAV while considering real-time need and distribution of ground users. In another study, the authors proposed a framework for UAV trajectory optimization for emergency response in post-disaster scenarios~\cite{requested_ref_1}. Their method utilizes intelligent connected groups of energy-efficient UAVs by employing deep reinforcement learning to address mobility control, limited energy capacity, and restricted communication ranges.

\subsection{UAV Placement Optimization}
Several studies have focused on the optimization of UAV position as mobile BSs in 5G networks, considering factors such as user coverage \cite{9838347}, backhaul connectivity \cite{10001472}, and user demand \cite{sun2022flexedge}. In \cite{al2020optimal}, the authors proposed an optimal 3D UAV base station placement approach that incorporates autonomous coverage hole detection, wireless backhaul, and user demand considerations. Their work addresses the challenge of ensuring seamless coverage while maximizing the utilization of UAVs.
In another study \cite{sabzehali2022optimizing}, the authors investigated the optimization of the number, placement, and backhaul connectivity of multi-UAV networks. Their research addresses the challenges associated with deploying multiple UAVs to achieve efficient network coverage while considering the limitations of backhaul connections. They propose a comprehensive optimization framework to address these challenges and enhance network performance. Wu et al. \cite{wu2023uavs} focused on UAV deployment algorithms specifically designed to maximize backhaul flow in 5G networks. Their work emphasizes the importance of efficient backhaul connectivity in UAV deployments and presents algorithms to enhance the utilization of UAVs while maximizing data transfer between the base stations and the core network. In \cite{farooq2018multi}, the researchers realized the dynamic aspects of connectivity in multi-UAV networks like the mobility of underlay devices and unavailability due to failures. They proposed a feedback based adaptive and resilient framework for dynamic deployment of UAVs as mobile BSs.

\begin{table*}[t]
\renewcommand{\arraystretch}{1.5}    \centering
    \begin{tabularx}{0.98\textwidth}{|c|ccc|ccc|ccc|c|}
        \cline{2-10}
         \multicolumn{1}{c|}{} & \multicolumn{3}{c|}{UAV Trajectory Design} & \multicolumn{3}{c|}{UAV Placement Optimization} & \multicolumn{3}{c|}{QoS Aware UAV Deployment}\\
        \cline{1-11}
        \multicolumn{1}{|c|}{\textbf{Features}} & \multicolumn{1}{c}{Refs. [25]} & \multicolumn{1}{c}{[26]} & \multicolumn{1}{c|}{[27]} & \multicolumn{1}{c}{Ref. [33]} & \multicolumn{1}{c}{Ref. [34]} & \multicolumn{1}{c|}{Ref. [35]} & \multicolumn{1}{c}{Ref. [38]} & \multicolumn{1}{c}{Ref. [39]} & \multicolumn{1}{c|}{Ref. [40]} & \multicolumn{1}{c|}{Our Solution} \\
        \cline{1-11}
        \textbf{\begin{tabular}[c]{@{}c@{}}Real-time Placement\end{tabular}} & $\times$ & $\checkmark$ & $\checkmark$ & $\checkmark$ & $\times$ & $\checkmark$ & $\times$ & $\times$ & $\checkmark$ & $\checkmark$ \\
        \cline{1-11}
        \textbf{\begin{tabular}[c]{@{}c@{}}Fronthaul-Backhaul\end{tabular}} & $\checkmark$ & $\checkmark$ & $\times$ & $\checkmark$ & $\checkmark$ & $\checkmark$ & $\checkmark$ & $\times$ & $\times$ & $\checkmark$ \\
        \cline{1-11}
        \textbf{\begin{tabular}[c]{@{}c@{}}Dynamic Nodes\end{tabular}} & $\times$ & $\checkmark$ & $\times$ & $\times$ & $\times$ & $\times$ & $\times$ & $\checkmark$ & $\checkmark$ & $\checkmark$ \\
        \cline{1-11}
        \textbf{\begin{tabular}[c]{@{}c@{}}Multiple UAVs\end{tabular}} & $\checkmark$ & $\checkmark$ & $\times$ & $\times$ & $\times$ & $\checkmark$ & $\times$ & $\checkmark$ & $\checkmark$ & $\checkmark$ \\
        \cline{1-11}
        \textbf{\begin{tabular}[c]{@{}c@{}}Deep Learning Approach\end{tabular}} & $\checkmark$ & $\times$ & $\times$ & $\times$ & $\checkmark$ & $\checkmark$ & $\times$ & $\checkmark$ & $\checkmark$ & $\checkmark$ \\
        \cline{1-11}
    \end{tabularx}
    \caption{A synopsis of recent works in UAV-assisted network optimization.}
    \label{table-contributions1}
\end{table*}

\subsection{QoS Aware UAV Deployment}
To ensure the QoS and reliability of multi-UAV networks, authors in \cite{hu2021uplink} propose an uplink throughput optimization scheme for UAV-enabled urban emergency communications. Their approach uses a joint power allocation and UAV altitude control scheme to maximize the uplink throughput of the UAV network. Meanwhile, other researchers consider the optimization of user coverage and throughput in multi-tier variable height UAV networks~\cite{nafees2021multi}. They propose a dynamic altitude adjustment scheme that optimizes the coverage and throughput of the UAV network. In another work, the authors introduce a DRL based framework specifically designed for the dispatch of UAVs in urban vehicular networks~\cite{requested_ref_2}. The study addresses challenges posed by intermittently connected vehicular networks, optimizing UAV deployment to ensure long-term communication relays with minimized energy consumption and efficient coverage. Several studies focus on resilient UAV orchestration and formation schemes for QoS-driven connectivity and coverage of ground users \cite{wang2022proactive,wang2022resilient}. These methods are based on dynamic UAV swarming and aim to maintain the connectivity and coverage of spatially dispersed ground users under various network failures and specific QoS requirements for users.

\subsection{Contribution}
Table~\ref{table-contributions1} provides a synopsis of recent works which can be broadly categorized into those considering real-time placement, fronthaul-backhaul tradeoff, dynamic nodes, multiple UAVs, and approaches using deep learning methodologies.
In this work, A novel DRL-aided UAV-assisted system is proposed for finding the optimal UAV placement for maximising the joint reward
function based on the uplink and downlink data rates. We address the critical coupling between fronthaul and backhaul links and leverage real-time performance measurement of users to optimize UAV placement for maximal coverage and enhanced overall system performance. The key contributions are summarized as follows:
\begin{itemize}
    \item We propose a real-time and dynamic placement approach for UAV placement in 5G networks. By integrating online learning techniques, the UAVs can efficiently adapt their positions to cater to changing user demands and network conditions, cater to the on-demand connectivity needs of critical users, and effectively enhancing overall performance.
    \item Unlike conventional approaches that often neglect the importance of backhaul links, this work explicitly considers the tradeoff between fronthaul and backhaul connectivity. By striking a well-balanced equilibrium between these links, the UAVs can ensure robust communication with ground users while maintaining seamless connectivity with the central infrastructure.
    \item Leveraging IAB, this work explores the potential of UAVs as dynamic and mobile DU nodes. The proposed approach showcases the efficacy of UAVs in augmenting coverage for short duration and effectively addresses coverage challenges in critical areas.
    \item In complex scenarios with a higher number of users and rapidly changing user locations, centralized optimization methods may prove limited. This work introduces a decentralized and adaptive DRL-based framework, distributing the burden among individual UAVs. This decentralization enhances the scalability and responsiveness of the system, making it suitable for large-scale networks with numerous users.
\end{itemize}

\begin{table}
\centering
\renewcommand{\arraystretch}{1.4}
\resizebox{\linewidth}{!}{
\begin{tabular}{|c|l|}
    \hline
    \textbf{Symbol} & \textbf{Description} \\
    \hline
    $\mathcal{U}=\{1,2,\ldots,U\}$ & Set of UAVs\\
    \hline
    $\mathcal{M}=\{1,2,\ldots,M\}$ & Set of ground users \\
    \hline
    $q_s\in\mathbb{R}^2$ & Coordinate of ground BS \\
    \hline
    $y_m\in\mathbb{R}^2,m\in \mathcal{M}$ & Coordinates of users\\
    \hline
    $q_u\in\mathbb{R}^3,u\in \mathcal{U}$ & Coordinates of UAVs \\
    \hline
    $D_{\text{s},m},m\in\mathcal{M}$ & Distance between ground BS and user\\
    \hline
    $D_{u,m},u\in\mathcal{U},m\in \mathcal{M}$ & Distance between UAV and user\\
    \hline
    $D_{\text{s},u},u\in\mathcal{U}$ & Distance between ground BS and UAV\\
    \hline
    $r$ & UAV BS communication range  \\
    \hline
    $p_{LoS},p_{NLoS}$ & Probability of line-of-sight/none-line-of-sight\\
    \hline
    $\vartheta,\xi$ & Environmental path-loss parameters \\
    \hline
    $\delta$ & Path-loss exponent\\
    \hline
    $\eta_{Los},\eta_{NLos}$ & Additional average path-loss \\
    \hline
    $B_u$ & Bandwidth of link between UAVs and users \\
    \hline
    $B_\text{s}$ & Bandwidth of link between ground BS and UAV \\
    \hline
    $\widetilde{\text{SNR}}_{\text{s},m}$ & SNR of direct link between BS and user \\
    \hline
    $\text{SNR}_{u,m}$ & SNR of fronthaul link between UAV and user \\
    \hline
    $\widehat{\text{SNR}}_{u,u'}$ & SNR of backhaul link between two UAVs \\
    \hline
    $\overline{\text{SNR}}_{\text{s},u}$ & SNR of backhaul link between BS and UAB \\
    \hline
    $\widetilde{C}_{\text{s},m}$ & Data rate of direct link between BS and user \\
    \hline
    $\text{C}_{u,m}$ & Data rate of fronthaul link between UAV and user \\
    \hline
    $\widehat{\text{C}}_{u,u'}$ & Data rate of backhaul link between two UAVs \\
    \hline
    $\overline{\text{C}}_{\text{s},u}$ & Data rate of backhaul link between BS and UAV \\
    \hline
    $\alpha$ & Reward balancing parameter\\
    \hline
    $\mu$ & Learning rate  \\
    \hline
    $\epsilon_{\max},\epsilon_{\min},\epsilon_{\Delta}$ & Max, min and decaying epsilon parameter  \\
    \hline
\end{tabular}}
\caption{Table of notations.}
\label{table symbol notation}
\end{table}

\section{System Model}
We consider a UAV-assisted cellular network as shown in Fig.~\ref{fig:system} which consists of a cellular BS located at $q_s=[x_s,y_s,z_s]^T\in \mathbb{R}^3$, a set of UAVs $\mathcal{U}=\{1,2,\ldots,U\}$ and a group of ground users $\mathcal{M}=\{1,2,\ldots,M\}$. The users are assumed to be located in a 2D plane and have high quality of service requirements \cite{MC_user1}.
The geographical locations of the users are denoted by $y=[y_1,y_2,...,y_M]^T$, where $y_m\in \mathbb{R}^2, \forall m \in \mathcal{M}$. The distance between the cellular BS and the user $m$ is denoted by $D_{\text{s},m}=\|q_s-y_m\|,\ m\in\mathcal{M}$. The users can connect with the cellular BS through direct link or using the UAV BS as a relay depending on the received QoS, e.g., higher signal to interference and noise ratio (SINR). 
A group of UAVs with mobile BSs are deployed to leverage IAB access and enhance the QoS of the users. It is assumed that all UAVs hover at the same height $h$ and their coordinates are represented by $q_u=[x_u,y_u,h]^T \in \mathbb{R}^3, u\in{\mathcal{U}}$. The maximum coverage range of the UAV is denoted by $r \in \mathcal{R}$, which implies that a user can be connected to the UAV only if the distance between them satisfies $D_{u,m}=\|q_u-y_m\| \leq r,\ m \in \mathcal{M}, u\in{\mathcal{U}}$.
The UAV also uses the access network to maintain backhaul connectivity with the cellular BS. Hence, the backhaul link exists if $D_{\text{s},u}=\|q_u - q_s \| \leq r$, and the ordering of the connected UAV network is denoted by $\Psi=\{\psi_1,\psi_2,\cdots,\psi_U\}$, which is a permutation of the set $\mathcal{U}$ such that $\psi_1$ refers to the UAV closest to the BS and $\psi_U$ refers to the UAV closest to the users.
The exact locations of the users are unknown to the UAV BS, however, their instantaneous QoS is known through the backhaul link with the cellular BS. Moreover, the users are assumed to be static or moving at low velocities compared with the UAV BS \cite{user_static}.
Since the main focus of this paper is the 3D placement of UAV, we neglect the velocity and energy consumption of UAV in the system modeling and algorithm design. In the following subsections, we describe the wireless channel models and the performance metrics used. 



\subsection{Fronthaul and Direct Link Path-Loss Model}

\begin{figure}[t]
    \centering
    \includegraphics[width=\linewidth]{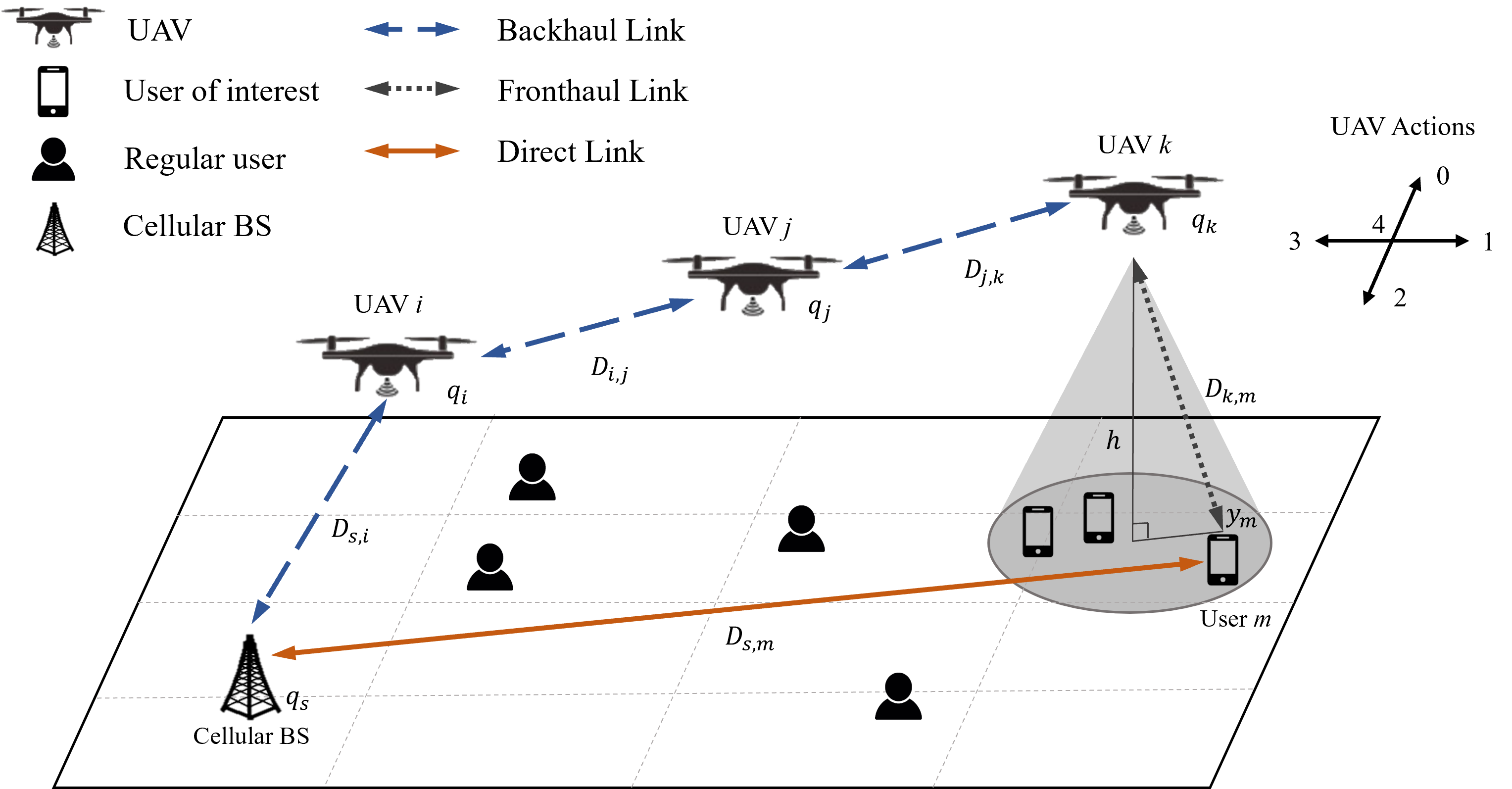}
    \caption{Illustration of system model. \vspace{-0.0in}}
    \label{fig:system}
\end{figure}

The UAV-to-user fronthaul links and the cellular BS-to-user direct links experience path loss in either a line-of-sight (LoS) path or a non-line-of-sight (NLoS) path, depending on various factors such as presence of obstacles and reflections from infrastructure, which can be modeled probabilistically. The probability of having a LoS path
is determined by~\cite{air-to-ground-pl}: \vspace{-0.0in}
\begin{equation}
    p_{\text{LoS}}=\frac{1}{1+\vartheta\exp\left(-\xi\frac{180}{\pi}\phi-\vartheta\right)},
\end{equation}
where $\vartheta$ and $\xi$ are constants depending on the environment and $\phi$ is the elevation angle as shown in Fig.~\ref{fig:system}. The probability of having an NLoS path is determined by $p_{\text{NLoS}}=1-p_{\text{LoS}}$. Each of these paths has a different signal attenuation based on the altitude, elevation angle and type of the propagation environment.
In general, NLoS connections tend to experience higher path loss compared to LoS connections. We can express the path loss between the UAV/cellular BS and user $m$ as a :
\vspace{-0.0in}
\begin{equation}
\begin{aligned}
\text{PL}_{k,m}=&10\log_{10}\left(\frac{4\pi f_c D_{k,m}}{c} \right)^\delta\hspace{-0.03in}+p_{\text{LoS}}\hspace{0.01in}\eta_{\text{LoS}} +
p_{\text{NLoS}}\hspace{0.01in}\eta_{\text{NLoS}},
\end{aligned}
\label{eq:pl_usr}
\end{equation}
where $D_{k,m}=D_{\text{s},m}$ when referring to the link between the user and the cellular BS, and $D_{k,m}=D_{u,m}$ when referring to the link between the user and the UAV. $c$ denotes the speed of light, $f_c$ is the carrier frequency, and $\delta$ is the path-loss exponent. The first term in \eqref{eq:pl_usr} represents the free-space path-loss, while $\eta_{\text{LoS}}$ and $\eta_{\text{NLoS}}$ are the additional average losses for LoS and NLoS paths respectively.

\subsection{Backhaul Path-Loss Model}
The backhaul link betweeen two UAVs or UAV and cellular BS are at relatively high altitude and can be considered to be similar to propagation in free space. Thus, we use a free space path loss model to describe the air-to-air (A2A) link path-loss between two UAVs $u$ and $u'$, or between UAV $u$ and cellular base station \cite{FSPL}. These can be expressed as follows:

\begin{equation}
    \begin{aligned}
    &\text{FSPL}_{u,u'}=20\log_{10}\left(\frac{4\pi f_a D_{u,u'}}{c}\right),\\
    &\text{FSPL}_{\text{s},u}=20\log_{10}\left(\frac{4\pi f_a D_{\text{s},u}}{c}\right),
\end{aligned}
\end{equation}
where $f_a$ is the frequency of the wireless channel used, $D_{u,u'}$ refers to the Euclidean distance between the UAV $u$ and UAV $u'$, and $D_{\text{s},u}$ refers to the Euclidean distance between the UAV $u$ and cellular base station.

\vspace{-0.0in}
\subsection{Quality of Service}
To quantify the QoS of ground users, we use the Shannon capacity based on the SNR. When the user $m$ is connected with the cellular BS through direct link, the SNR of the link can be expressed as:
\vspace{-0.0in}
\begin{equation}
    \widetilde{\text{SNR}}_{\text{s},m}=\frac{P_r^{\text{s},m}}{\sigma^2},\hspace{-0.0in}
\label{eq:direct sinr}
\end{equation}
where $P_r^{\text{s},m}=P_t^{\text{s},m}/10^{\text{PL}_{\text{s},m}/10}$ is the received signal power by the user, $P_t^{\text{s},m}$ is the transmission power of the UAV and $\sigma^2$ is the thermal noise power. $P_r^{\text{s},m},P_t^{\text{s},m}$ and $\sigma^2$ are all in the units of mW. Consequently, the achieved data rate can be calculated using the bandwidth of the direct link $B_{\text{s}}$:
\vspace{-0.0in}
\begin{equation}
    \widetilde{C}_{\text{s},m}=B_{\text{s}}\hspace{0.02 in}\log_2(1+\text{SNR}_{\text{s},m}).\vspace{-0.0in}
\label{eq:direct_c}
\end{equation}
When the user is connected with a UAV, the the fronthaul link SNR of user $m$ connected by UAV can be expressed as:
\vspace{-0.0in}
\begin{equation}
    \text{SNR}_{u,m}=\frac{P_r^{u,m}}{\sigma^2},\hspace{-0.0in}
\label{eq:sinr}
\end{equation}
where $P_r^{u,m}=P_t^{u,m}/10^{\text{PL}_{u,m}/10}$ is the received signal power by the user and $P_t^{u,m}$ is the transmission power of the UAV. The achieved data rate of the fronthaul link depends on the available bandwidth of the wireless link between the UAV $u$ and the users, denoted by $B_u$, and can be expressed as:
\vspace{-0.0in}
\begin{equation}
    C_{u,m}=B_u\hspace{0.02 in}\log_2(1+\text{SNR}_{u,m}).\vspace{-0.0in}
\label{eq:fronthaul_c}
\end{equation}

The SNR for the backhaul links can be defined for two possible connection types: 1) between two UAVs $u$ and $u'$; and 2) between the UAV $u$ and cellular BS. As both connections utilize free-space path-loss model, the SNR is calculated by:
\vspace{-0.0in}
\begin{equation}
    \widehat{\text{SNR}}_{u,u'}=\frac{P_r^{u,u'}}{\sigma^2}, \hspace{0.2in}\overline{\text{SNR}}_{\text{s},u}=\frac{P_r^{\text{s},u}}{\sigma^2}, \vspace{-0.0in}
\end{equation}
where $P_r^{u,u'}=P_t^{u,u'}/10^{\text{FSPL}_{u,u'}/10}$ defines the received signal power of UAV $u$ and $P_t^{u,u'}$ refers to the transmission power of the UAV $u'$. $P_r^{\text{s},u}=P_t^{\text{s},u}/10^{\text{FSPL}_{\text{s},u}/10}$ defines the received signal power of backhaul link between the UAV and the cellular BS. Given the available bandwidth of the backhaul link with the cellular base station $B_{\text{s}}$ and the UAV $B_u$, the achieved data rate of the link can be expressed by:
\vspace{-0.0in}
\begin{equation}
\begin{aligned}
   &\widehat{C}_{u,u'}=B_{\text{s}}\hspace{0.02 in}\log_2(1+\widehat{\text{SNR}}_{u,u'}),\vspace{-0.0in}\\
   &\overline{C}_{\text{s},u}=B_{\text{s}}\hspace{0.02 in}\log_2(1+\overline{\text{SNR}}_{\text{s},u}),\vspace{-0.0in}
\end{aligned}
\label{eq:backhaul_c}
\end{equation}
As the users have the connectivity options to select between the direct link with the cellular BS or the fronthaul link with the UAV, given the connected UAV network $\Psi=\{\psi_1,\psi_2,\ldots,\psi_U\}$, the resulting effective data rate of user $m$ can be expressed as:
\vspace{-0.0in}
\begin{equation}
\begin{aligned}
    C_{m}=\max \big( \min &\big(\overline{C}_{\text{s},\psi_1},\widehat{C}_{\psi_1,\psi_2},\ldots,\\
    &\widehat{C}_{\psi_{U-1},\psi_U},C_{\psi_U,m} \big),\widetilde{C}_{\text{s},m}\big),\vspace{-0.0in}
    \label{eq:final snr}
\end{aligned}
\end{equation}
where $\overline{C}_{\text{s},\psi_1},\widehat{C}_{\psi_1,\psi_2},\ldots,\widehat{C}_{\psi_{U-1},\psi_U}$ represents the backhaul links between the UAVs and the cellular BS, and $C_{\psi_U,m}$ represents the fronthaul link connecting the UAV $U$ to the user $m$. The term $\min \big(\overline{C}_{\text{s},\psi_1},\widehat{C}_{\psi_1,\psi_2},\ldots,\widehat{C}_{\psi_{U-1},\psi_U},C_{\psi_U,m} \big)$ reflects the minimum data rate among the fronthaul and backhaul links and is the overall QoS evaluation of the UAV network. Depending on the maximum QoS, users have the flexibility to either connect with the UAV network or establish a direct link with the cellular BS.

\section{Methodology}
In this section, we formulate the UAV placement optimization problem and introduce our solution based on dueling double deep Q-network (D3QN), a widely-used reinforcement learning algorithm, to learn the optimal placement of UAVs for maximum data rates or QoS for users in 5G networks.
\subsection{Problem Formulation}
The goal of the optimization problem is to find the optimal UAV position $q_u$ to maximize the sum of data rates of users defined in equation \eqref{eq:final snr} while satisfying the SNR threshold in equation \eqref{eq:reward}. The optimization problem can be formulated as follows:
\begin{subequations}
\begin{align}
    &\max_{q_u}\sum_{m\in \mathcal{M}}C_{m},\\
    &\text{s.t.}\notag \\
    &\max ( \min (\overline{\text{SNR}}_{\text{s},\psi_1},\widehat{\text{SNR}}_{\psi_1,\psi_2},\ldots,\widehat{\text{SNR}}_{\psi_{U-1},\psi_U},\nonumber\\
    &\hspace{0.6in}\text{SNR}_{\psi_U,m}),\widetilde{\text{SNR}}_{\text{s},m})\geq T_1,\forall m\in\mathcal{M},\label{eq:con2}\\
    &\|q_s-q_{\psi_1}\|\leq r\hspace{0.05 in}, \label{eq:con3}\\
    &\|q_u-q_{u+1}\|\leq r,\hspace{0.05 in} \forall u\in\Psi \backslash \psi_U, \label{eq:con4}\\
    &0\leq x_u \leq x_{\max}, 0\leq y_u \leq y_{\max}, u\in\mathcal{U}, \label{eq:con6}
\end{align}
\end{subequations}
where 
constraint \eqref{eq:con2} ensures that the effective end-to-end SNR is above a  threshold;
\eqref{eq:con3} ensures that the cellular base station is connected to the closest UAV;
\eqref{eq:con4}  represents the backhaul connectivity constraint to form a connected UAV network;
\eqref{eq:con6} limit the movement boundaries for the UAV.
The described problem is a nonlinear and non-convex optimization problem since it contains nonlinear and non-convex inequality constraints. Existing solutions to such problems quantize them into several convex subproblems or solve the transformed problem with relaxed constraints. However, the optimized results are only applicable to the current environment and may fail or lose accuracy in complex situation. The global optimal solution may not exist, e.g., dispersed users. However, local optimum may be possible. Our proposed method is based on Q-learning which learns the environment from continuous observations and value iterations. The UAV placement algorithm is designed such that it can balance the fronthaul and backhaul links, adapt to complex environment and obtain a feasible solution to the above problem.

\subsection{Preliminaries}
In this section, we introduce the fundamental concept of Q-learning. The Q-learning algorithm is a model-free, iterative method that uses a Q-function to estimate the expected reward of taking a specific action in a particular state. At each iteration, the agent selects an action according to the policy, and receives a reward based on the selected action and the current state. For an agent at state $s$ that performs action $a$ by the policy $\pi$, the value function is defined by the expected cumulative reward:
\begin{equation}
\label{eq:q value}
    V(s,a)=\mathbb{E}\left[\sum_t \gamma R(s_t,a_t)|s_1=s,a_1=a,\pi\right]
\end{equation}
where $\gamma\in (0,1)$ is the discount factor, $R(s_t,a_t)=\sum_t \gamma^{t-1} r(s_t,a_t)$ is the cumulative discounted reward given $s$, $a$ and $\pi$. In finite iterations, there is always an optimal policy $\pi^*$ where the Bellman optimality equation for the optimal state-action value function $Q^*(s,a)$ is given by:
\begin{equation}
\label{eq:bellman}
\begin{aligned}
    Q^*&(s,a)=\mathbb{E}\left[R(s,a)\right]+\gamma \sum_{s'\in S}P_{ss'}(a,\pi^*)V(s',\pi^*) \\
    &=\mathbb{E}\left[R(s,a)+\gamma \max_{a'\in \mathcal{A}} Q^*(s',a')\right],
\end{aligned}
\end{equation}
where the optimal polity $\pi^*$ is determined by choosing the action with maximum $Q^*$ value. The Q-function is updated iteratively based on the agent's experiences, with the goal of maximizing the cumulative reward over time. Based on \eqref{eq:bellman}, the Q-value is updated as follows:
\begin{equation}
\label{eq: q update}
\begin{aligned}
    Q(s,a)=&(1-\mu)Q(s,a)+\\
    &\mu\left[r(s,a)+\gamma\max_a Q(s',a')\right],
\end{aligned}
\end{equation}
where $\mu\in(0,1)$ is the learning rate.

In RL algorithms, it is challenging to balance between exploitation of the learned Q-function and exploration of new states. One common approach for selecting the action is the $\epsilon-$greedy policy defined by:
\begin{equation}
    a=\left\{
    \begin{aligned}
        &\arg\max_a Q(s,a)  & \text{w.p. }\epsilon,\\
        &\text{random action from }\mathcal{A}     & \text{w.p. }1-\epsilon,
    \end{aligned}\right.
\end{equation}
where $\epsilon\in(0,1)$ is the probability of selecting optimal action.

\subsubsection{Observation Space and Action Space}
We begin by defining the observation and action space for the UAV. Assume that the UAV is not aware of the coordinates of the ground users. Instead, at each timeslot $t$, it can observe the achieved data rates of users using the backhaul link. The lists of achieved data rates is represented by $\mathcal{C}(t)=[C_1(t),C_2(t),...,C_m(t)], m\in \mathcal{M}$.
\begin{itemize}
    \item \textbf{State space:} The state of the UAV at timeslot $t$ is denoted as $s_t=[x_u^t,y_u^t,h]$, where $(x_u^t,y_u^t,h)$ is the 3-D coordinates of the UAV, which are bounded by $0\leq x_u^t\leq x_{\max}$, $0\leq y_u^t\leq y_{\max}$. 
    \item \textbf{Action space:} The action space is shown in Fig. \ref{fig:system}. At each state $s_t$, the UAV can select an action $a_t$ out of five candidate options represented by $\mathcal{A}\in\{0,1,2,3,4\}$ and the state is updated by $s_{t+1}=s_t+a_t\in \mathcal{A}|s_t$. Actions ``0" to ``3" move the UAV one step North, East, South, and West respectively. Action ``4" corresponds to the UAV hovers in the current position.
\end{itemize}


\subsubsection{Dueling Double Deep Q Network}



\begin{figure}[t]
    \centering
    \includegraphics[width=\linewidth]{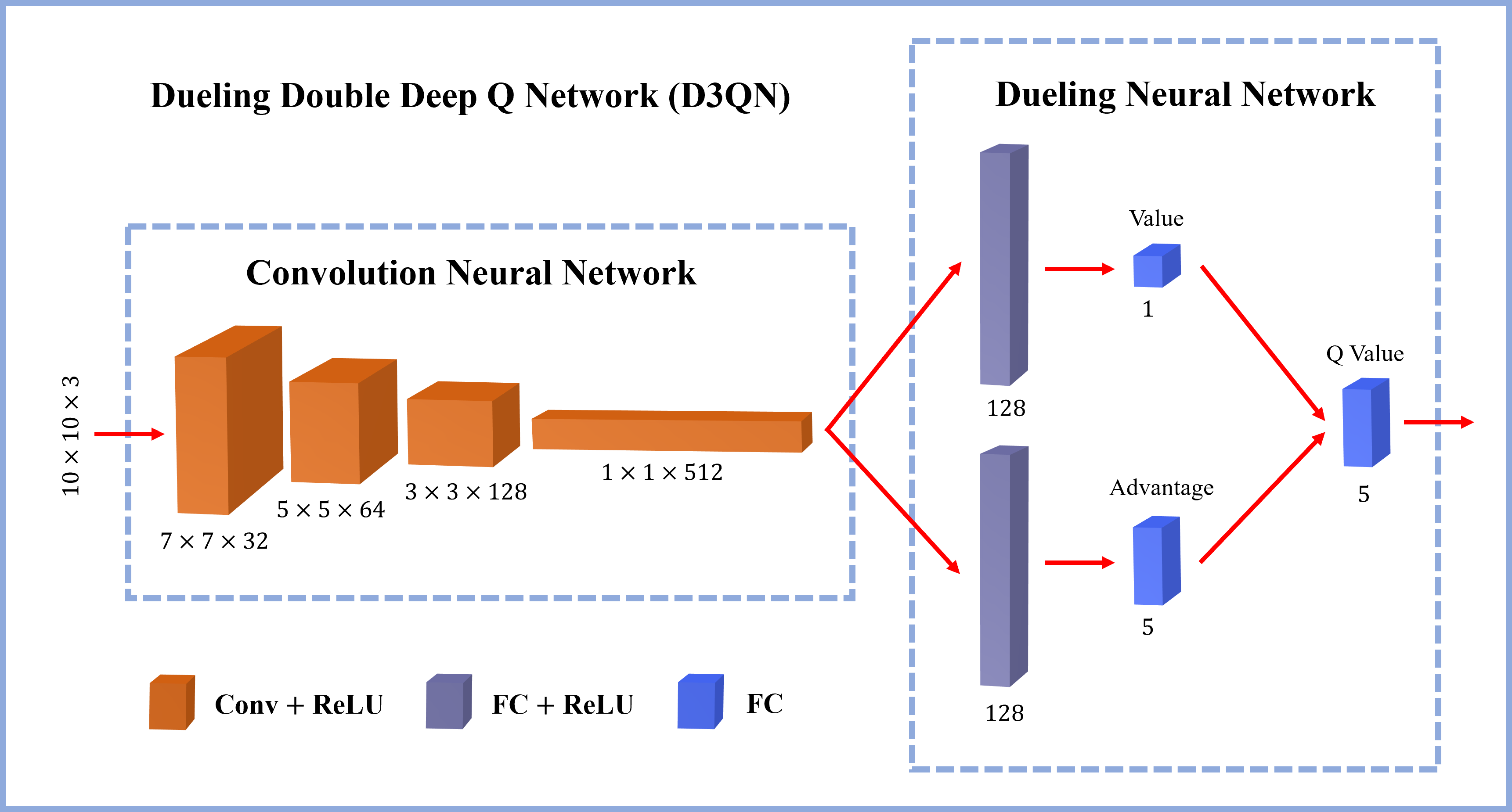}
    \caption{The architecture of the D3QN model. The input of the network is the 3D coordinates of the UAVs. It is followed by four convolutional layers and two fully connected layers for two streams of dueling structure.}
    \label{fig:D3QN}
\end{figure}

D3QN is an extension of the deep Q-Network algorithm that integrates experience replay, dueling architecture and double learning network to enhance stability and performance. The structure of the D3QN model is shown in Fig. \ref{fig:D3QN}. In a dueling DQN, the states are first fed to a convolutional feature learning network, followed by the dueling architecture which separates the two steams of state values and action advantages \cite{yan2020towards}. The two steams are then combined by a fully connected layer to generate an estimate of the Q function. This separation enables the network to focus on learning the quality of actions independently of the state, thereby augmenting the agent's proficiency in understanding complex action relationships. The double learning network employ two distinct Q networks with one for active learning and the other as a target network for Q value estimating \cite{zeng2021simultaneous}. This dual-network approach contributes to stabilizing the learning process and mitigating the risk of overestimating Q values.

\subsubsection{Reward Function}
The goal of the reward function is to deliver a robust and consistent signal for the users. In the design of the the reward function, we include both mean and $75^{th}$ percentile data rates to ensure fair data rates for all users while also addressing the needs of users with higher demands. The incorporation of mean data rates ensures equitable data rates and promotes fair user experience. By considering the $75^{th}$ percentile data rates, the network guarantees a certain level of service for a significant portion of users. With the recorded user data rates $\mathcal{C}(t)$, we design the reward function using the mean $\Bar{\mathcal{C}}(t)$ and $75^{th}$ percentile $\hat{\mathcal{C}}(t)$ of the data rates distribution as follows:\vspace{-0.0in}
\begin{equation}
\label{eq:reward}
    r(s_t,a_t)=
        \alpha\Bar{\mathcal{C}}(t)+(1-\alpha)\hat{\mathcal{C}}(t),
\end{equation}
where $\alpha$ is the reward balancing parameter.

\subsection{D3QN Based UAV Placement}
In the design of the D3QN based UAV placement algorithm, we used an decaying $\epsilon$-greedy exploration strategy to balance between exploration and exploitation. We started with a high $\epsilon$ value to encourage exploration at the beginning of the learning process, and gradually reduced it as the agent gained more experience and knowledge. Specifically, we used a linear decay schedule, where $\epsilon$ was decreased linearly from an initial value of $\epsilon_{\max}=0.99$ to a final value of $\epsilon_{\min}=0.1$ over a pre-defined number of iterations. This allowed the agent to explore the state-action space in the beginning, and gradually focus on exploiting the learned knowledge as the training progressed. We chose the decay schedule based on our prior experience and empirical evaluations. By decreasing the exploration rate over time, we ensured that the agent focused more on exploiting the learned knowledge as the training progressed, and avoided getting stuck in sub-optimal policies due to excessive exploration.

\vspace{-0.0in}
\begin{algorithm}
    \caption{\textbf{Q-Learning Based UAV Placement}}
    \label{alg:UAV placement}
\begin{algorithmic}[1]
\State \textbf{Initialize:} 
\State Sets of ground users $\mathcal{M}$;
\State UAV initial position and initial state~$q_u(0)=s_0 \gets [0,0,h]$;
\State Initial ordering of the UAV network $\Psi=\{1,2,\cdots,U\}$;
\State Initialize Q-values: $Q(s,a)=0,\ \forall s,a$;
\State Set initial exploration rate: $\epsilon\gets \epsilon_{\max}$;
\State Set current episode: $e=0$, total episodes: $e_{\max}=100$; 
\State Set current iteration: $i \gets 0$, total iterations: $i_{\max} \gets 100$;
\While{$e<e_{\max}$}
    \State Increment episode: $e\gets e+1$
    \While{$i<i_{\max}$}
        \State Increment iteration: $i\gets i+1$
        \State Compute $\text{SNR}_{u,m},\text{SNR}_{\text{s},u},\text{SNR}_{\text{s},m}$
        \If{ $\min \left(\text{SNR}_{u,m},\text{SNR}_{\text{s},u}\right)>\text{SNR}_{\text{s},m}$}
            \State User $m$ connects to UAV.
        \Else
            \State User $m$ connects to cellular BS.
        \EndIf
        \State Update ordering of the UAV network $\Psi$ using the sequence of the connected UAV nodes from Algorithm \ref{alg:UAV formation}.
        \State Update $\mathcal{C}(t)$, $\Bar{\mathcal{C}}(t)$ and $\hat{\mathcal{C}}(t)$.
        \State Generate random sample $\Lambda$ from $\text{Uniform}(0,1)$.
        \If {$\Lambda<\epsilon$}
            \State{UAV selects a random action $a_t$ from action space $\mathcal{A}$ uniformly}.
        \Else
            \State{Select action $a_t$ that maximizes future reward: $a_t\gets \max_{a'}Q(s_{t+1},a')|s_t$.}
        \EndIf
        \State Reduce exploration rate: $\epsilon\gets \max(\epsilon_{\min},\epsilon-\epsilon_{\Delta})$
        \State Update state: $s_{t+1}\gets s_t+a_t$.
        \State Update Q-values using equation \eqref{eq: q update}.
    \EndWhile
\EndWhile
\end{algorithmic}
\end{algorithm}

\begin{figure}[t]
    \centering
    \includegraphics[width=\linewidth]{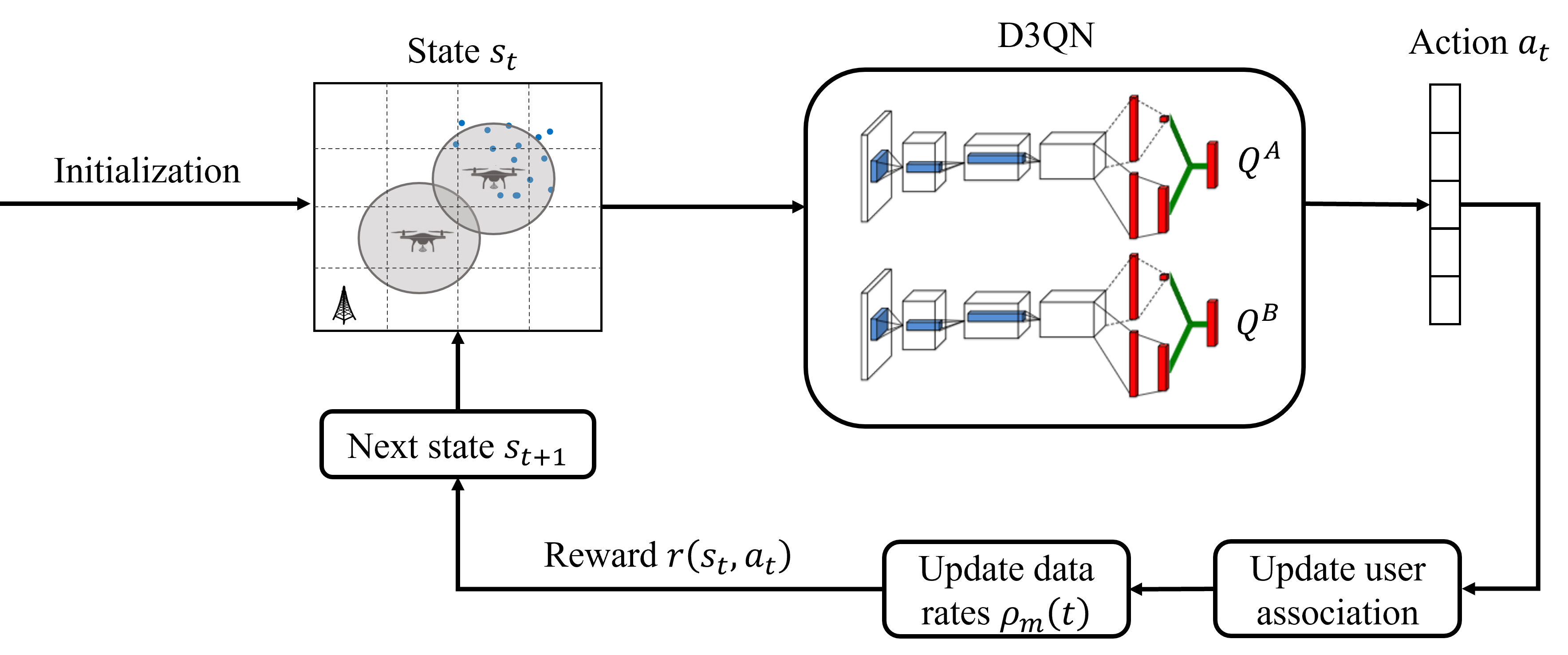}
    \caption{The flowchart of the proposed algorithm.}
    \label{fig:algorithm}
\end{figure}

The D3QN based UAV placement algorithm is presented as Algorithm \ref{alg:UAV placement}. At the beginning of the task, the algorithm randomly initializes the coordinates of the ground users using a 2D normal distribution and the initial position of the UAV BS is set to $[0,0,h_1]$. The Q value table is initialized with zeros for each state-action pair. Parameters like learning rate $\mu$, discount factor $\gamma$, exploration factor $\epsilon$ and its decaying rate are given in Table \ref{table1}. In each episode of the task, the UAV starts from the initial position and traverse the grids by selecting action $a_t\in\mathcal{A}$ based on the $\epsilon$ greedy exploration policy. At each iteration, the values of $\text{SNR}_{u,m},\text{SNR}_{\text{s},u},\text{SNR}_{\text{s},m}$ are updated and used to decide whether the user is connected by the UAV or the cellular BS. With the UAV positions and the UAV to user association, the backhaul network formation algorithm utilizes breadth-first-search (BSF) to explore the possible connections between neighboring UAVs satisfying the certain SNR threshold, and generates the sequence of the backhaul link as in Algorithm \ref{alg:UAV formation}. The UAV receives the list of achieved data rates of users $\mathcal{C}(t)$ from the cellular BS and receives a reward based on the mean data rates $\Bar{\mathcal{C}}(t)$ and $75^{th}$ percentile data rates $\hat{\mathcal{C}}(t)$. The Q table is then updated using equation \eqref{eq: q update}. At last, the $\epsilon$ value is decayed and the position of the UAV BS is updated. The process is updated until the maximum training iterations are reached and the algorithm converges to an optimal policy that achieve the maximum reward in the shortest path. The algorithm flow is illustrated in Fig. \ref{fig:algorithm}. 

The computational complexity of the proposed algorithm can be determined based on three main components, i.e., the processing time of the deep learning output, the fronthaul user association time, and the backhaul formation time. Firstly, the deep learning model processing time refers to the duration it takes for the D3QN model to process a UAV state input and generate the probability array of actions. This duration is determined by the model architecture and hardware specifications but can generally be considered as a constant. Secondly, the fronthaul association involves assessing all possible combinations of users and UAVs, resulting in a time complexity of $\mathcal{O}(M\cdot U)$. Thirdly, for determining the backhaul connectivity network, we employ an algorithm based on BFS, outlined in Algorithm \ref{alg:UAV formation}. The time complexity of this algorithm is dependent on the number of UAVs and is expressed as $\mathcal{O}(U^2)$. In summary, the overall time complexity of the proposed UAV placement algorithm can be expressed as $\mathcal{O}(M\cdot U + U^2)$, reflecting the combined computational requirements of fronthaul association and backhaul formation.


\begin{algorithm}
    \caption{Backhaul Formation Algorithm}
    \label{alg:UAV formation}
    \begin{algorithmic}[1]
    \Initialize{Sets of ground users $\mathcal{M}$;\\
    Sets of UAVs $\mathcal{U}$;\\
    Ground BS $\mathcal{S}$;\\
    Sets of explored nodes $l_x=\{\mathcal{S}\}$;\\
    Sets of unexplored nodes $l_u=\{1,2,\ldots,\mathcal{U}\}$.
    }
    \Input{
    Coordinates of the UAVs $q_u,u\in\mathcal{U}$;\\
    Coordinates of the ground BS $q_s$;\\
    Maximum communication range $r$.}
    \While{$l_u$ is not empty}
        \State Get the first unexplored node from $l_u$
        \For{$j\textbf{ in }l_x$}
        \State Update $\text{SNR}_{i,j},\text{SNR}_{j,s}$
        \If{ $\text{SNR}_{i,j},>\text{SNR}_{j,s}$ and $\text{SNR}_{i,j}>T_1$}
        \State UAV $j$ is connected by UAV $i$.
        \State Add $j$ to explored nodes $l_x$.
        \State Break \textbf{for} loop.
        \EndIf
        \EndFor
        \State Assign the sequence of connected UAVs $\Psi \gets l_x$.
    \EndWhile
\end{algorithmic} 
\end{algorithm}

\section{Simulation Results}
\begin{table}
\begin{tabularx}{0.48\textwidth}{|X|c|}
    \hline
    \textbf{Parameter} & \textbf{Value} \\
    \hline
    Number of ground users, $M$ & 100 \\
    \hline
    UAV BS communication range, $r$ & 500 m \\
    \hline
    Bandwidth of link between UAVs and users, $B_u$ & 25 MHz\\
    \hline
    Bandwidth of link between cellular BS and UAV, $B_{\text{s}}$ & 25 MHz\\
    \hline
    Environmental path-loss parameters, $\vartheta,\xi$ & 4.88, 0.43\\
    \hline
    Path-loss exponent, $\delta$ & 2.0\\
    \hline
    Additional average LoS path-loss, $\eta_{Los}$ & 0.1 dB\\
    \hline
    Additional average NLoS path-loss, $\eta_{NLos}$ & 21.0 dB\\
    \hline
    Reward balancing parameter, $\alpha$ & 0.5\\
    \hline
    Learning rate, $\mu$ & 0.01 \\
    \hline
    Discount factor, $\gamma$ & 0.9 \\
    \hline
    Maximum, minimum and decaying epsilon greedy parameters, $\epsilon_{\max},\epsilon_{\min},\epsilon_{\Delta}$ & 0.99, 0.01, 0.01 \\
    \hline
\end{tabularx}
\caption{Parameter values used in simulations. }
\label{table1}
\end{table}
In this section, we present the simulation results to demonstrate the performance of the proposed D3QN based UAV placement algorithm.

\subsection{Experimental Setup}
We perform a series of simulations in Python 3.9 environment using TensorFlow 2.11.0. Unless otherwise stated, we consider a simulation area of size 1 km $\times$ 1 km with a cellular base station located at $[10,0,10]$ and $M=100$ ground users. The users are distributed according to 2D Gaussian distribution with a mean $[70,70]$ and a covariance matrix $[[100,0],[0,50]]$, which is a realistic user distribution in post-disaster or remote areas. We deploy 3 UAVs with their initial positions set to $[0,0,h]$. The communication range and the altitude for the UAVs are set to $r=500\ $m and $h=100~\text{m}$, which are designed for operations in suburban areas as per our prior work \cite{yuhi_ieee_mass}. The bandwidth of the link between UAVs and users is set to be equivalent as the link between cellular BS and UAV for the simplicity of simulation. The chosen set of environmental path-loss parameters align with prevalent parameters utilized in similar studies. The path-loss exponent is set to $\delta=2.0$ which adheres to the standard free-space propagation model \cite{al2020modeling}. In order to simulate the characteristics of a contemporary cellular network environment, the additional average LoS and NLoS path-loss follow the study in \cite{al2020optimal} and are tailored to $\eta_{LoS}=0.1$ dB and $\eta_{NLoS}=21.0$ dB respetively. The learning rate and discount factor are tuned to $\mu=0.01$ and $\gamma=0.9$ during the training of the D3QN model, aiming to optimize the trade-off between swift convergence and stability, as well as temporal rewards against long-term gains \cite{9739716}. As for the epsilon-greedy exploration parameters, they are designed to achieve a balance between exploration and exploitation within the state space \cite{ouahouah2021deep}. The main parameters of the simulation setting are listed in Table \ref{table1}.

\subsection{Performance Evaluation}

\begin{figure}[t]
    \centering
    \includegraphics[trim={0 0 0 0.4in},clip,width=\linewidth]{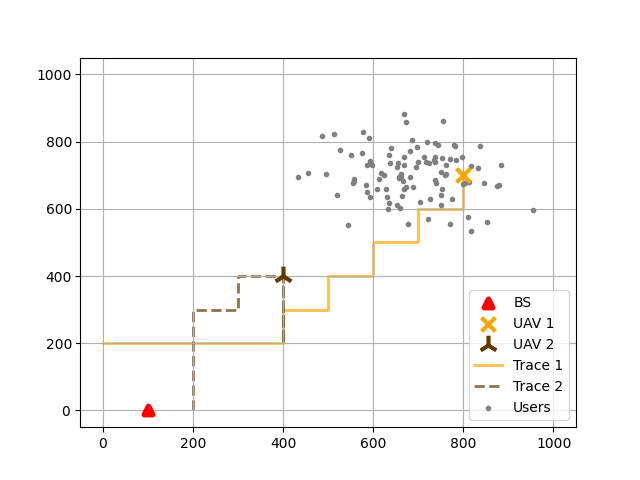}
    \caption{Top view of the UAV placements and trajectories after training convergence with two UAVs. One UAV serve as the access point for the users, while the other UAV provides backhaul connection with the cellular BS. \vspace{-0.0in}}
    \label{fig:res1}
\end{figure}

\begin{figure}[t]
    \centering
    \includegraphics[trim={0 0 0 0.4in},clip,width=\linewidth]{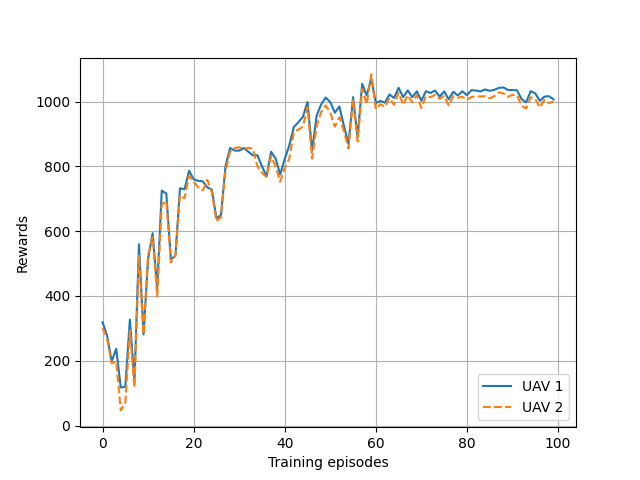}
    \caption{Training reward against number of episodes with two UAVs. The training reward converges after 60 episodes and reach an optimal value of around 1000. \vspace{-0.0in}}
    \label{fig:res6}
\end{figure}

In Fig.~\ref{fig:res1}, we show the top view of the learned trajectories of the UAV BSs and the distribution of ground users. Red triangle denotes the cellular BS, grey dots denote the ground users, the yellow line and brown dashed line denote the learned trajectories. This simulation uses a UAV step size of $10 \text{m}$ and SNR threshold $T_1=3.0$. The two UAVs started from $[20,0]$ and $[0,20]$ respectively and formed a connected backhaul network between the cellular BS and the users. The UAV 1 provides direct connection with the users and learned its optimal position at $[80,70]$, while UAV 2 serves as a relay node and learned its optimal position at $[40,40]$. Fig.~\ref{fig:res6} shows the curves of training rewards against training episodes for the two UAVs respectively. It can be shown that two UAVs constantly learn and update their action policy to enhance the data rates of the users.

\begin{figure}[t]
    \centering
    \includegraphics[trim={0 0 0 0.4in},clip,width=\linewidth]{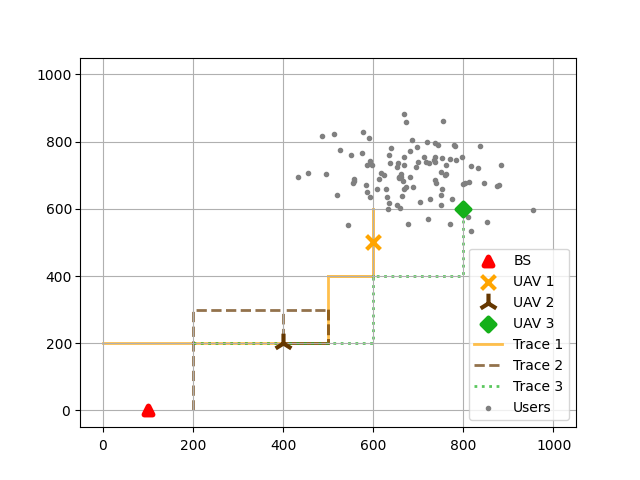}
    \caption{Top view of the UAV placements and trajectories after training convergence with three UAVs. One UAV serve as the access point for the users, while the other two UAVs provides backhaul connection with the cellular BS. \vspace{-0.0in}}
    \label{fig:res3}
\end{figure}

\begin{figure}[t]
   \centering
   \includegraphics[trim={0 0 0 0.4in},clip,width=\linewidth]{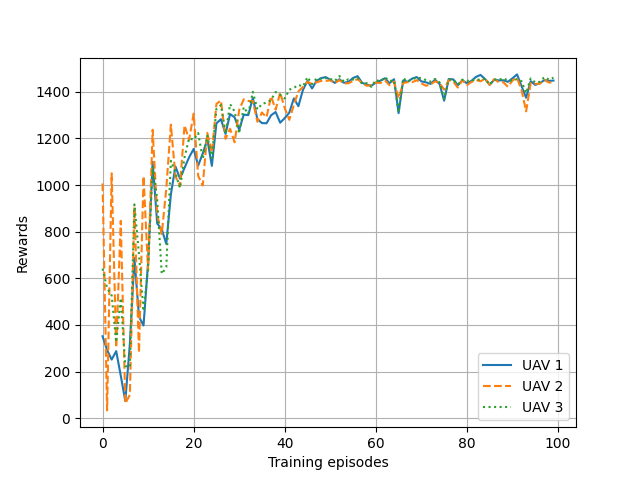}
   \caption{Training reward against training episodes with three UAVs. The training reward converges after 50 episodes and reach an optimal value of around 1400.\vspace{-0.0in}}
   \label{fig:res9}
\end{figure}

In Fig.~\ref{fig:res3}, we increase the number of UAVs to $M=3$ and show the top view of their learned trajectories. This simulation uses the same UAV step size of $10 \text{m}$ and SNR threshold $T_1=3.0$. In contrast to the result in Fig.~\ref{fig:res1} which used 2 UAVs, the learned optimal positions of UAV 1,2 and 3 are at $[60,50]$, $[40,20]$, $[80,60]$ respectively. UAV 2 serves as a relay node to link the cellular BS and UAV 1. UAV 1 links UAV 3 and UAV 2, meanwhile provides direct connections to users with UAV 3 based on the SINR selection. Fig.~\ref{fig:res9} shows the curves of training rewards against training episodes for the three UAVs respectively. The results demonstrate that the three UAVs can form a connected UAV network to enhance signal strength and increase coverage for the users. 


\begin{figure}[t]
    \centering
    \includegraphics[trim={0 0 0 0.4in},clip,width=\linewidth]{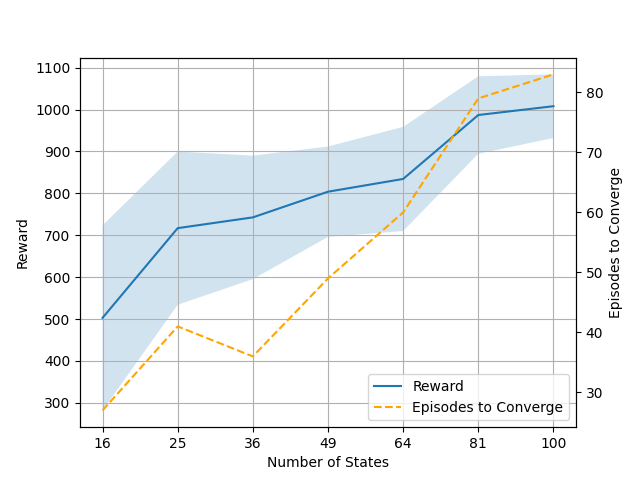}
    \caption{The trade-off between the training rewards and the training costs with respect to the number of states. Both the training rewards and the episodes to convergence increase as the number of states increase.\vspace{-0.0in}}
    \label{fig:res10}
\end{figure}

In Fig. \ref{fig:res10}, we explore the influence of the number of state variables on the model performance and training costs. The ground plane is divided into grids with varying sizes 
$\{4\times4,5\times5,6\times6,7\times7,8\times8,9\times9,10\times10\}$ 
representing possible candidate UAV locations. When the number of states is less than 81, the performance increases at decaying speed. In contrast, when the number of states is larger than 81, the performance reaches maximum value and the reward curve keeps oscillating. This result shows the tradeoff between performance and costs in the training process: as the number of states increases, the training rewards after convergence increases at decaying rate until some maximum values. However, the training episodes before convergence also increase exponentially.

\begin{figure}[t]
    \centering
    \includegraphics[trim={0 0 0 0.4in},clip,width=\linewidth]{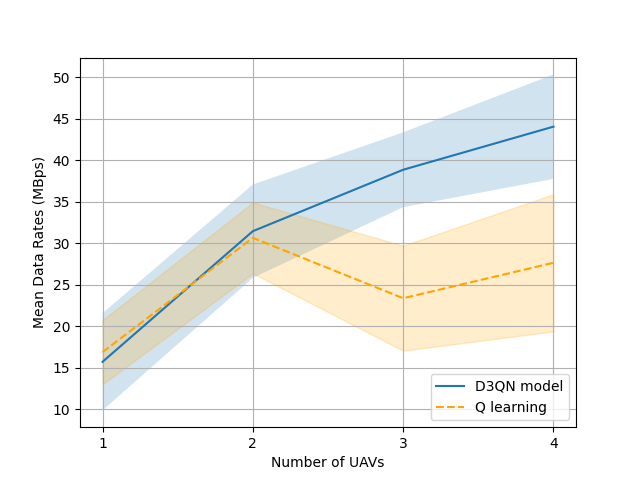}
    \caption{Performance comparison of the proposed method with the baseline method using different number of UAVs. Our proposed method achieves better mean data rates in complex situations with more UAVs.\vspace{-0.0in}}
    \label{fig:res11}
\end{figure}


To highlight the performance of the proposed learning based algorithm, we use the model presented in \cite{yuhi_ieee_mass} as baseline which assumed the weighted geographic center as the placement goal. In Fig.~\ref{fig:res11}, we compare the performance of the proposed model with the baseline model with respect to the number of UAVs. In less complex situations when the number of users is less than 2, the performance difference between the proposed model and the baseline model is negligible. Both of them can converge well and achieve optimal user data rates. However, in complex environment with large number of UAVs, the proposed model outperforms the baseline model because the state space expand exponentially with the number of UAVs and the Q-learning model is unable to learn and converge in complex situations. In contrast, the proposed D3QN model is based on deep neural networks and incorporates the dueling double network features, which enables the method to learn the state-value function efficiently and reduce the risk of overestimation of the Q-values.

\begin{figure}[t]
    \centering
    \includegraphics[trim={0 0 0 0.4in},clip,width=\linewidth]{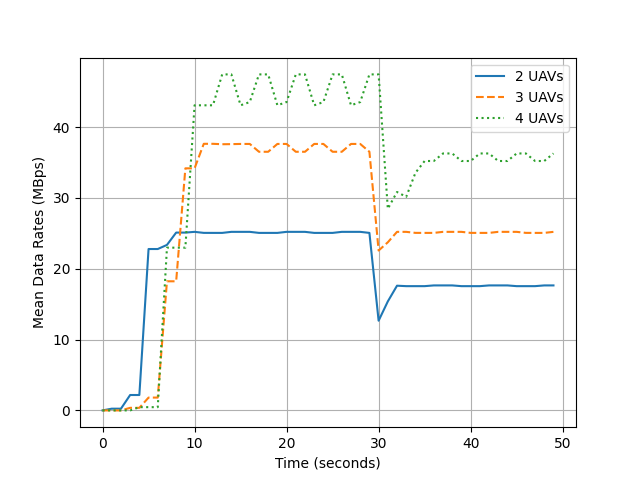}
    \caption{Experiment on the resilience of the UAV network. One of the UAVs is selected randomly to fail at $t=30s$. The UAVs can adopt pretrained models and recover the formation rapidly.\vspace{-0.0in}}
    \label{fig:res12}
\end{figure}
\begin{figure}[t]
    \centering
    \includegraphics[trim={0 0 0 0.4in},clip,width=\linewidth]{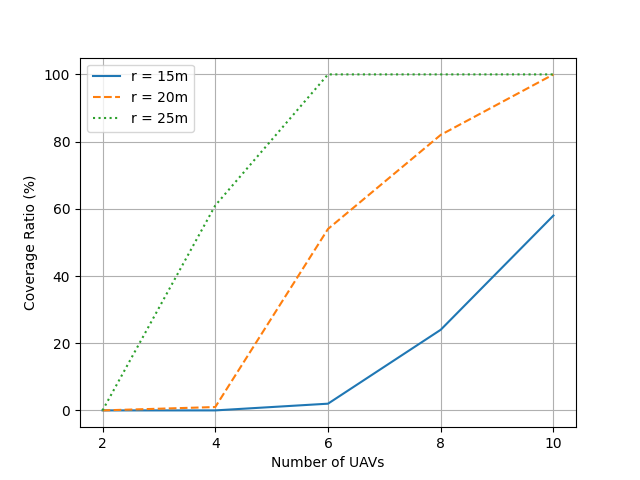}
    \caption{Comparison of user coverage ratio with respect to the number of UAVs using different communication range $r$. \vspace{-0.0in}}
    \label{fig:res13}
\end{figure}
\begin{figure}[t]
    \centering
    \includegraphics[trim={0 0 0 0.4in},clip,width=\linewidth]{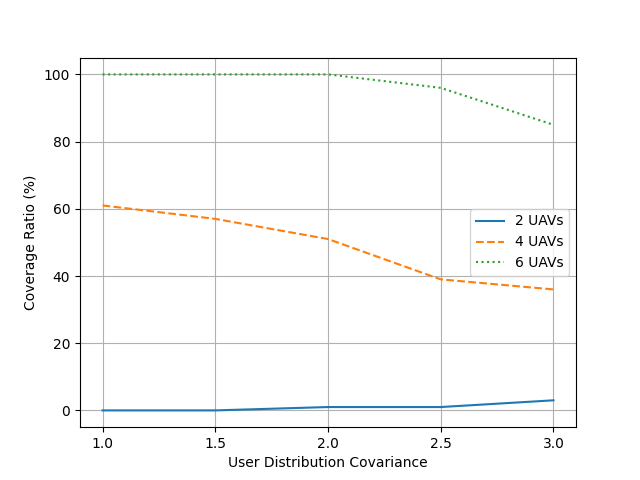}
    \caption{Comparison of user coverage ratio with respect to the user distribution covariance using different number of UAVs. The base distribution covariance is set to [[100,0],[0,50]].\vspace{-0.0in}}
    \label{fig:res14}
\end{figure}

The Fig. \ref{fig:res12} illustrates the resilience of the UAV network through a series of experiments lasting 50 seconds each, involving 2, 3 and 4 UAVs respectively. In each experiment, a UAV is randomly selected to fail at $t=30 s$, simulating a real-world scenario of network disruptions. Remarkably, these UAVs are equipped with pretrained models, allowing for rapid recovery and formation reconstruction. In the case of utilizing 2 UAVs, the system converges efficiently, achieving a maximum mean data rate of 25 MBps. After the unexpected failure, it swiftly recovers, reconstructs its formation, and reaches a commendable maximum mean data rate of 18 MBps. The scenario with 3 UAVs exhibits even higher data rates, peaking at 38 MBps during convergence. Following the random failure and subsequent recovery, the system effectively reconstructs and reattains a data rate of 25 MBps. Lastly, with 4 UAVs in action, the system reaches a substantial maximum mean data rate of 44 MBps during convergence. Even after encountering a random failure and recovery process, it demonstrates remarkable resilience by reconstructing and achieving a robust maximum mean data rate of 38 MBps. These findings address the robustness and adaptability of the proposed UAV placement algorithm, even in the face of unforeseen disruptions, showcasing its potential for reliable data transmission and communication.

In Fig. \ref{fig:res13}, we presents a comprehensive analysis of user coverage ratio concerning the number of UAVs, with three distinct experiments utilizing varying UAV communication ranges of $r=15, 20$ and $25$ m. In the case of $r=15$ m, the coverage ratio remains stagnant at 0\% when only 2 and 4 UAVs are deployed, indicating that the UAV network isn't sufficiently proximate to the users in these scenarios. However, as the number of UAVs surpasses 6, the coverage ratio begins to escalate, ultimately reaching a peak of 48\%. Moving to the scenario with $r=20$ m, the coverage ratio exhibits a similar behavior of remaining at 0\% when only 2 UAVs are deployed. As the number of UAVs increases beyond 2, the coverage ratio experiences growth, albeit at a decreasing rate. When 10 UAVs are deployed, the coverage ratio attains a full 100\%. Finally, in the case of $r=25$ m, a rapid increase in coverage ratio to 100\% occurs when the number of UAVs reaches 6, demonstrating the remarkable effectiveness of this configuration in ensuring user coverage. 

Fig. \ref{fig:res14} illustrates the dynamic relationship between user coverage ratio and the covariance of user distribution. We conducted three experiments employing 2, 4, and 6 UAVs under a UAV communication range of $r=30$ m, and a base user distribution covariance of $[[100,0],[0,50]]$. In the initial case involving 2 UAVs, the coverage ratio remains at 0\% for covariance of 1.0 and 1.5 times the base value, indicating that the UAV network is not sufficiently close to the users in these scenarios. However, as the covariance increases to 2.0 times the base value, users start to spread out, leading to a coverage ratio peak of 3\% at a covariance of 3.0. In the scenario with 4 UAVs, the coverage ratio peaks at a substantial 61\% when the covariance is 1.0, subsequently decreasing gradually to a minimum of 36\% at a covariance value of 3.0. Remarkably, when utilizing 6 UAVs, the coverage ratio maintains a constant 100\% for covariances of 1.0, 1.5, and 2.0 times the base value, emphasizing the network's ability to provide comprehensive user coverage. However, as the covariance increases to 3.0, the coverage ratio gradually drops to a minimum of 85\%, indicating a reduction in user coverage as users disperse further. These results show the critical impact of user distribution characteristics on UAV network performance and can aid in optimizing UAV deployment strategies for diverse covariance scenarios.

\section{Conclusion}

In this paper, we have presented a dynamic, real-time approach for optimizing end-to-end performance in UAV-assisted wireless networks. Leveraging integrated access and backhaul capabilities, the proposed method utilizes deep reinforcement learning to efficiently orchestrate UAVs for service delivery to ground users in various scenarios such as remote areas, disaster-struck regions, and mission-critical operations.
Unlike traditional strategies that often overlook the vital role of backhaul links, our approach seeks a balanced optimization between fronthaul and backhaul connections. This balance is critical to ensuring a robust network formation, thereby achieving high-quality service for end-users. The reward function, based on fronthaul and backhaul data rates, was employed to guide the learning process, driving the system towards an optimal state that maximizes user data rates while maintaining strong backhaul connectivity among UAVs.
Our evaluation across different scenarios corroborates the effectiveness of the proposed approach. The results demonstrate that the system can autonomously place UAVs and efficiently balance the trade-offs between backhaul and fronthaul links. This results in optimized IAB performance, offering enhanced coverage and data rates to users on the ground.  
Nevertheless, there are several limitations in the application of the proposed solution in real-world environments. The collection of training data may be restricted due to security or privacy concerns, and the training process may suffer from battry constraints if the training is done using real UAVs. Although the proposed framework exhibits faster convergence compared to traditional Q learning, it still necessitates substantial training for effective operation. Future work may extend this framework to incorporate more complex conditions such as dynamic user mobility, energy-efficient UAV operation, and distributed learning.

\ifCLASSOPTIONcaptionsoff
  \newpage
\fi

\bibliographystyle{IEEEtran}
\bibliography{IEEEabrv,Bibliography}

\begin{IEEEbiography}[{\includegraphics[width=1in,height=1.25in,clip,keepaspectratio]{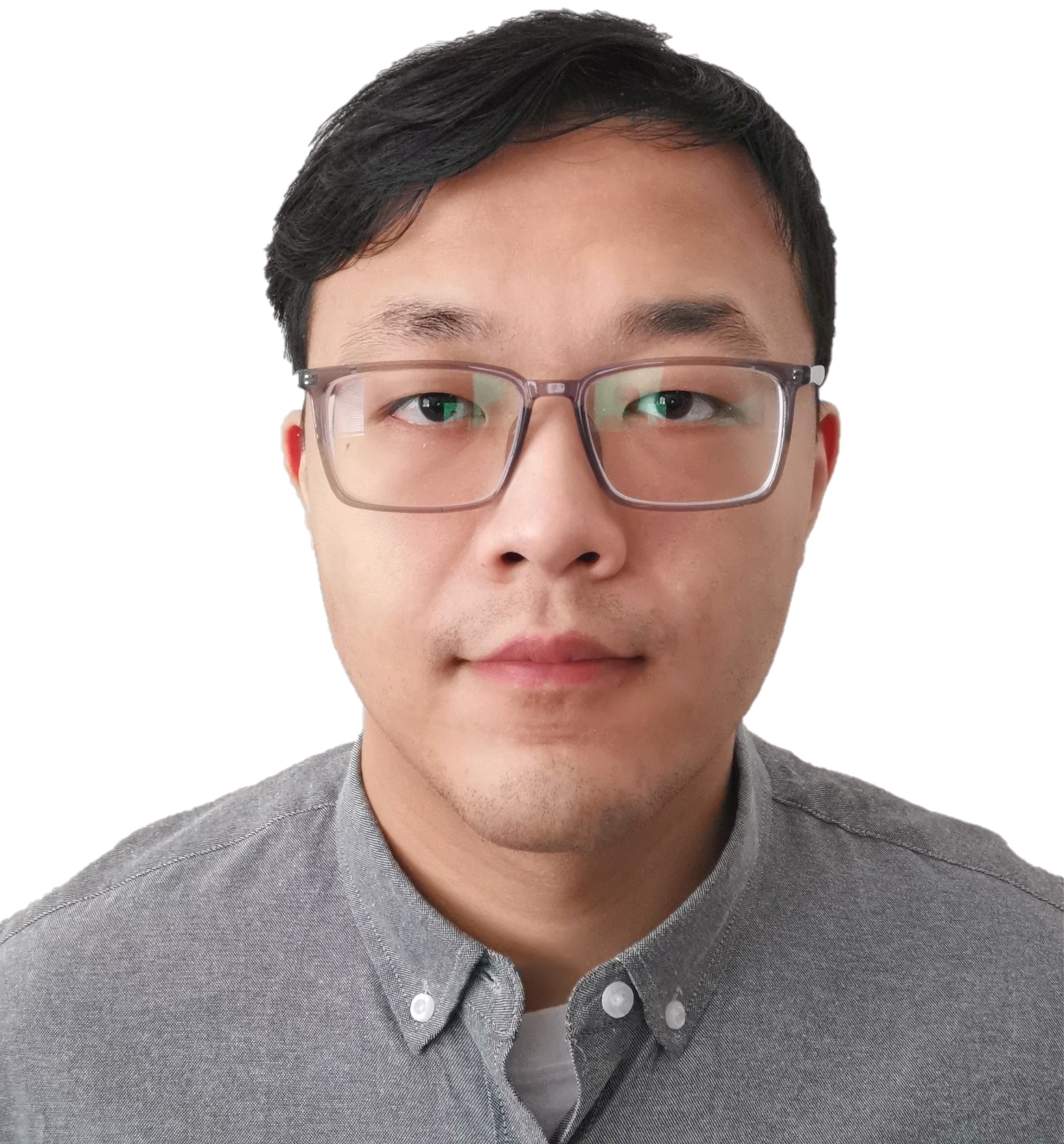}}]{Yuhui Wang}
(S'20) received the B.Eng. degree in Computer Science from Hong Kong University of Science and Technology (HKUST), Hong Kong, China in 2019 and the M.S. degree in Computer Science from New York University (NYU), Brooklyn, NY, in 2021. He is currently working toward the Ph.D. degree with the department of Electrical and Computer Engineering, University of Michigan-Dearborn. His research interests include mobile edge computing, machine learning, UAV networks and Internet of Things.
\end{IEEEbiography}

\begin{IEEEbiography}[{\includegraphics[width=1in,height=1.25in,clip,keepaspectratio]{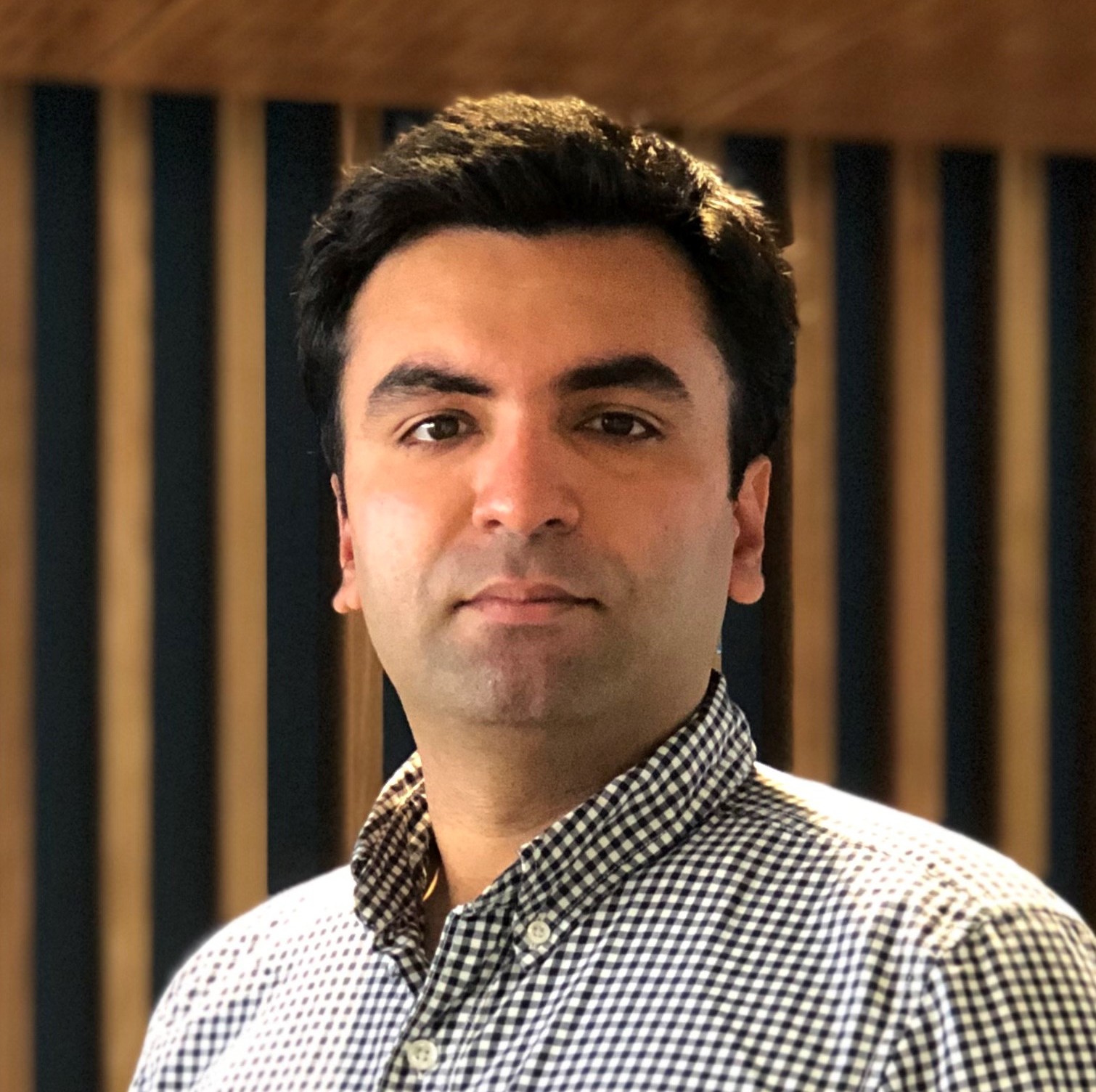}}]{Junaid Farooq}
(S'15-M'21) received the B.S. degree in electrical engineering from the School of Electrical Engineering and Computer Science (SEECS), National University of Sciences and Technology (NUST), Islamabad, Pakistan in 2013, the M.S. degree in electrical engineering from the King Abdullah University of Science and Technology (KAUST), Thuwal, Saudi Arabia in 2015, and the Ph.D. degree in electrical engineering from the Tandon School of Engineering, New York University, Brooklyn, NY in 2020. He was the recipient of the NYU University wide Outstanding Dissertation Award in 2021. Currently, he is an Assistant Professor with the Department of Electrical and Computer Engineering, University of Michigan-Dearborn. His research interests include optimization, security, and resilience of communication networks, cyber-physical systems, and the Internet of things. 
\end{IEEEbiography}

\end{document}